# Machine Vision and Deep Learning for Classification of Radio SETI Signals

G. R. Harp[1], Jon Richards[1], Seth Shostak[1] Jill C. Tarter[1]
Graham Mackintosh[2], Jeffrey D. Scargle[3], Chris Henze[3], Bron Nelson[3],
G. A. Cox[4], S. Egly[5], S. Vinodababu[5], J. Voien[5]

[1] *The SETI Institute, 189 Bernardo Ave., Mountain View, CA, 94043; gerryharp@gmail.com*
[2] *IBM Emerging Technology; gmackint@us.ibm.com*
[3] *NASA Ames Res. Center, Center for Supercomputing, Moffett Field, CA, USA*
[4] *IBM Watson Data Platform; adamcox@us.ibm.com*
[5] *Team Effsubsee; stephane.egly@gmail.com, voien21@gmail.com*

## ABSTRACT

*We apply classical machine vision and machine deep learning methods to prototype signal classifiers for the search for extraterrestrial intelligence. Our novel approach uses two-dimensional spectrograms of measured and simulated radio signals bearing the imprint of a technological origin. The studies are performed using archived narrow-band signal data captured from real-time SETI observations with the Allen Telescope Array and a set of digitally simulated signals designed to mimic real observed signals. By treating the 2D spectrogram as an image, we show that high quality parametric and non-parametric classifiers based on automated visual analysis can achieve high levels of discrimination and accuracy, as well as low false-positive rates. The (real) archived data were subjected to numerous feature-extraction algorithms based on the vertical and horizontal image moments and Huff transforms to simulate feature rotation. The most successful algorithm used a two-step process where the image was first filtered with a rotation, scale and shift-invariant affine transform followed by a simple correlation with a previously defined set of labeled prototype examples. The real data often contained multiple signals and signal ghosts, so we performed our non-parametric evaluation using a simpler and more controlled dataset produced by simulation of complex-valued voltage data with properties similar to the observed prototypes. The most successful non-parametric classifier employed a wide residual (convolutional) neural network based on pre-existing classifiers in current use for object detection in ordinary photographs. These results are relevant to a wide variety of research domains that already employ spectrogram analysis from time-domain astronomy to observations of earthquakes to animal vocalization analysis.*

Keywords: A list of 3–5 keywords are to be supplied.

## 1 Introduction

Machine learning or deep learning (ML/DL) and convolutional neural networks (C/NN) have recently achieved astonishing success for automated image classification and object identification. Such non-parametric techniques are said to have the potential to revolutionize science, especially with research involving very large datasets. Unfortunately, it is often difficult to adapt deep learning strategies to scientific problems when the format of the science data is not homomorphic with one of a small number of representations supported by canned deep learning solutions. For example, complex and large sequences of time series (voltage) data from a radio telescope might be considered analogous to natural language (Qing 2002) with e.g. recurrent neural networks applied to sampled acoustic time series (Sak et al. 2014). But the astronomical time series have thousands of times more time samples than an acoustic signal so natural language processing does not scale well to SETI.

For image processing, convolutional neural networks have shown rapidly improving outcomes for image parsing and classification, starting from AlexNet (Krizhevsky A et al. 2012), VGG (Simonyan & Zisserman 2015), and Inception (Szegedy et al. 2015) to deep Residual networks (He et al. 2016) and more recently, Wide Residual Networks (Zagoruyko & Komodakis 2016a, 2016b) which have the extra advantage of being computationally more efficient.

In this paper we explore some machine vision algorithms for classifying SETI signals using voltage time series converted into two-dimensional spectrogram "images." By reshaping the linear time series into a 2D format we open the door to application of well-known machine vision algorithms, including CNNs. The results are very promising and we consider how a ML classifier could be integrated into the near real-time SETI signal detector (SonATA) at the Allen Telescope Array.

The application of machine vision algorithms to spectrograms is clearly an idea whose time has come. Many ideas related to the use of convolutional neural networks to SETI spectrograms originated in the open-source contest submissions to a hackathon and code challenge (Cox & Harp 2017) and the source code from winning projects can be found at https://github.com/setiQuest/ML4SETI. Indeed, in what, for us, is the high signal to noise regime, a closely related work was recently reported (Tandiya et al. 2018) for radio frequency anomaly detection in wireless networks. In the low signal to noise regime, neural networks are being applied to identify fast radio bursts in simulations (Rankawat et al. 2018) and, quite impressively, in real observations (Zhang et al. 2018a). In what is actually an extension of this work, the detection of multiple (independent) signals in SETI spectrograms has been reported in (Rankawat & Harp 2018).

## 2.1   1D to 2D

To benefit from advances in ML, we pre-process long time series of (complex-valued, 8-bit real, 8-bit imaginary) voltage data by reshaping them from 1D to 2D to simulate an image. Domain knowledge and care are necessary in defining the 1D to 2D mapping, with a simple example in Figure 1. Here a synthetic voltage time series containing a substantial signal is mapped to a 2D figure in two different ways (time series, spectrogram). Firstly, the time series is broken into 128 equal length blocks representing rasters in the image. The rasters are then stacked in time order from bottom to top, creating a "waterfall" representation. In Figure 1 A, the squared magnitude (a.k.a. power) of each time series is plotted to form an image. This simple transformation does not bring out the structure of the weak signal and the resultant image appears to contain only noise.

The same processing is applied in Figure 1 B, except that each time series block is passed through a frequency filter bank based on a fast Fourier transform. In Figure 1 B, the signal power is concentrated into a small number of pixels in the image, allowing ready identification of the signal.

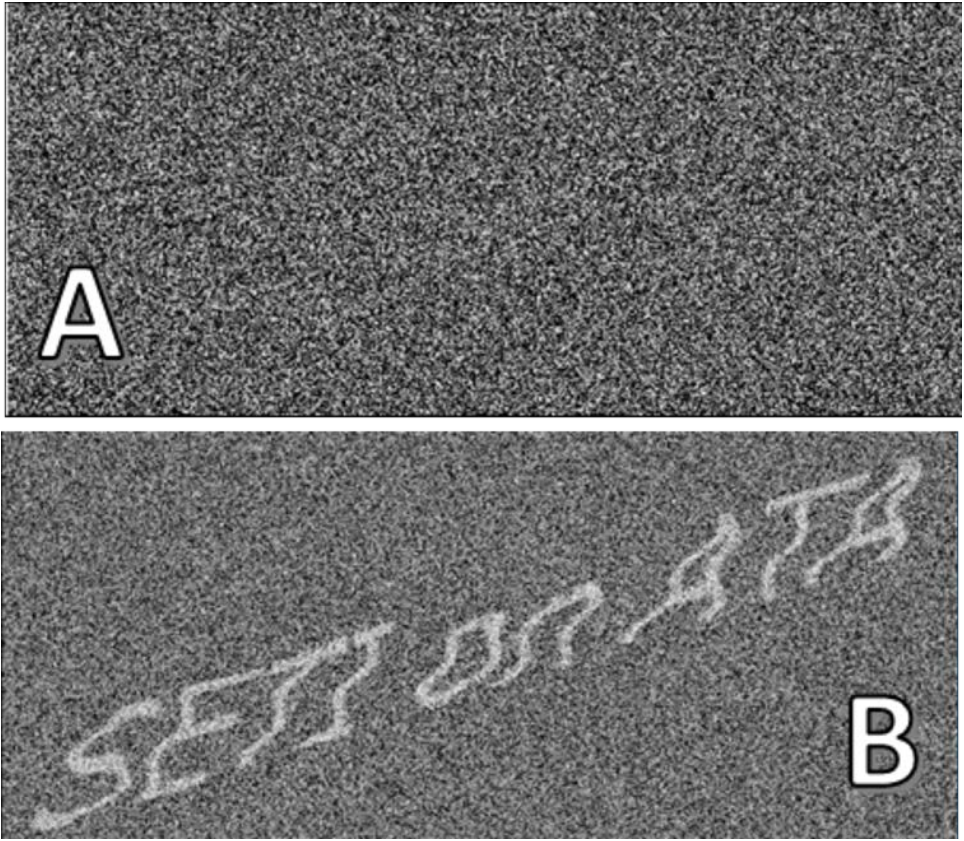

**Figure 1: A synthetic time series representing a radio signal is rendered as an image in two ways. A) The time series is split into 128 blocks of equal size, and signal power versus time in each block is represented as a grayscale raster (darker is lower power). B) The same data, after passing each raster line through at windowed Fourier transform. The Fourier transform is just one of a semi-infinite set of orthogonal transforms that might be used. Different kinds of transforms will be more or less sensitive to different kinds of signals.**

The simulated signal in Figure 1 is, of course, contrived. However, if we had no experience analyzing this type of data, then use of a frequency filter bank might not be an obvious step. Effectively, there are an infinite number of orthogonal linear transformations that might be used to prepare the data for imaging, and other choices may be reasonable (Harp et al. 2018). For example, variations of principle component decomposition (a.k.a. Karhunen–Loève Transform, a.k.a singular value decomposition) (Maccone 2010) are sometimes used to analyze time series raster data. Such decomposition results in an optimally compact representation of continuous signals and deduces the most important signal waveform(s) embedded in the signal. Yet principle component decomposition does not yield a 2D representation well suited to analysis with machine vision.

Figure 2 shows representative spectrograms selected from our database. The raw data are complex voltage values in a time series, approximately 70,000 complex values (sample rate = 760 samples per second over 92 seconds). The time series are broken up into 128 rasters of equal length (with 50% time overlap from one to the next) and lined up vertically. After a poly-phase filter bank, we selected a narrow range of frequencies (250 Hz) that contain representative signals. Each raster, with overlaps, spans approximately 1.44 seconds which is the effective integration time per sample in the spectrogram.

The definition of a 2D spectrogram is ambiguous with respect to the aspect ratio of the output image. For example, the data in a square spectrogram (equal number of discrete points on frequency and time axes) may be rendered as a rectangle by decreasing the point spacing on one axis while increasing the point spacing on the other axis by the same factor. Ideally, the aspect ratio would be adjusted to maximize the discrimination power between signal categories based on a sampling of real data. Our experiments indicate that a small change in aspect ratio (e.g. by a factor of 2) does not have a strong impact on the categorization results. In all cases here, the aspect ratio was set to some convenient value near (1:1), and this parameter was not optimized.

## 2.2 Signal to Noise Ratio

Deep learning has been very successful with image segmentation questions such as, given a photograph of a scene, does the scene contain a dog? In most applications, there is an implicit assumption that the photograph has a very high signal to noise ratio: that the value of (almost) every pixel contains highly relevant information about the scene. Conversely, astronomical images or spectrograms are often dominated by noise, and the relevant signal contained in a single pixel may be smaller than the noise.

This has important impacts on the applicability of machine vision to the problem of categorizing spectrograms. To a first approximation, every spectrogram looks like noise. But noise, by definition, is random so it is also true that every noise-dominated image is uncorrelated with every other such image at the pixel level. This rules out similarity or pattern matching methods that rely on naïve cross-correlation of spectrograms with a set of standard spectrograms.

In astronomy, low SNR can be often be mitigated by integrating a long time over many measurements. Because integration destroys time-dependent information, we choose to integrate over time periods where the desired signal is almost stationary; shorter or of order the "coherence time" for the measurement under investigation.

The signal detector at the Allen Telescope Array (ATA) is called SonATA (Harp et al. 2016) uses this approach by projecting the raw voltage data onto a small number of basis functions (drifting sinusoids) chosen in advance. A family of test signals is parameterized with one or more variables that vary over specific ranges. Then the observed signal is correlated with each test signal (matching filter for that test signal) and the correlation coefficient is recorded. Because a matched filter provides optimal sensitivity for the test signal, this method has close to optimal sensitivity for discovering signals with waveforms that are identical to the test signals (apart from a global phase). This is the state of the art detector in narrowband SETI searches though it has some disadvantages in terms of flexibility, e.g. (Harp et al. 2018).

Machine vision methods such as CNNs are infinitely more flexible in terms of signal waveforms that can be detected; the universal approximation theorem (Hornik 1991) states that any NN with three or more layers can be sized to approximate any function on a finite domain, even discontinuous functions. In effect, the NN automatically integrates over (ignores) all input parameters with low predictive power. Given a large database of training data examples, the NN, in some sense, chooses the optimal set of basis functions for discriminating data at hand.

Returning to the set of test functions prescribed in our SonATA detector, SonATA can identify features that are invisible to the eye (below the noise floor) in a naïve spectrogram. We expect that a machine vision approach to signal detection might have difficulty distinguishing very weak signals from pure noise. For the prototype study reported here, a rule of thumb is that if a human cannot identify the signal in the presence of noise, then a machine vision algorithm may completely miss the signal. This is a serious issue if machine vision methods are to replace traditional searches for SETI signals.

In cases where the classifying algorithm operates on an entire spectrogram at once (e.g. a CNN), there is no reason in theory that it could not achieve sensitivity comparable with our SonATA detector. We speculate that the limited sensitivity observed here could be overcome with the use of a much larger training set that densely samples the full range of signal to noise ratios that could theoretically be detected. In the prototype studies described here this requirement for training data is not met, thus our machine vision methods are effective only in cases of relatively high SNR.

## 2.3  Initial Attempts at De-noising

In a closely related result, we report that we have tried many kinds of convolutional noise-reduction at the spectrogram-pixel level, but no method has provided a convincing benefit. For example, low-pass, high-pass, and bandpass filters in the frequency domain[1] fail because the desired signal is distributed almost uniformly in frequency space. More generally, we understand that convolutional de-noising using small convolution kernels fails because the algorithms must be able to distinguish the signal from the noise using only a fraction of the pixels in the image, before that signal can be enhanced.

---

[1] Recall that any convolutional approach to data smoothing is equivalent to a filtering in frequency space.

One non-linear noise reduction algorithm leverages segmentation of the spectrogram using a watershed algorithm, e.g. (Gonzalez & Woods 2002; Isenstein & Hut 1998). The spectrogram is viewed as a height image, and water dropped at any pixel will move toward some local minimum or basin. The boundary between basins is defined by pixels where dropped water has equal probability of moving toward two adjacent basins.

If the spectrogram values are inverted (high power represented as black) before computing the watershed, each "basin" represents a bright feature or signal. The integrated power in each basin is computed and those with power below a threshold are discarded (flattened). In limited testing this watershed algorithm applied to spectrograms prior to signal detection gave a modest benefit for noise reduction. More work is required to make a quantitative decision about the benefits of watershed processing, and we suggest that this may be an interesting avenue for future study.

## 2.4  Large Database of Observed Candidate Signals

Through eight years of observations at the ATA, recordings of about 100,000,000 "candidate" signals have been made (candidates defined below). These are a filtered subset of all the artificial signals detected in our observations. Preliminary signal detection is performed by specialized software that identifies signals with bandwidth < 100 Hz hence are generated by technology (Grimm et al. 1987), in the midst of dominating noise (Harp et al. 2016). The noise usually dominates the signal, which is partly a selection effect. If there were a strong continuous signal from outer space that is easily detected, then by now it would have been found in other research. As such, most strong signals are associated with radio frequency interference. Although our system does not reject strong signals until they are characterized, most of the signals considered interesting (candidate signals) are close to the lower threshold for signal power specified in the SonATA detector (Harp et al. 2016).

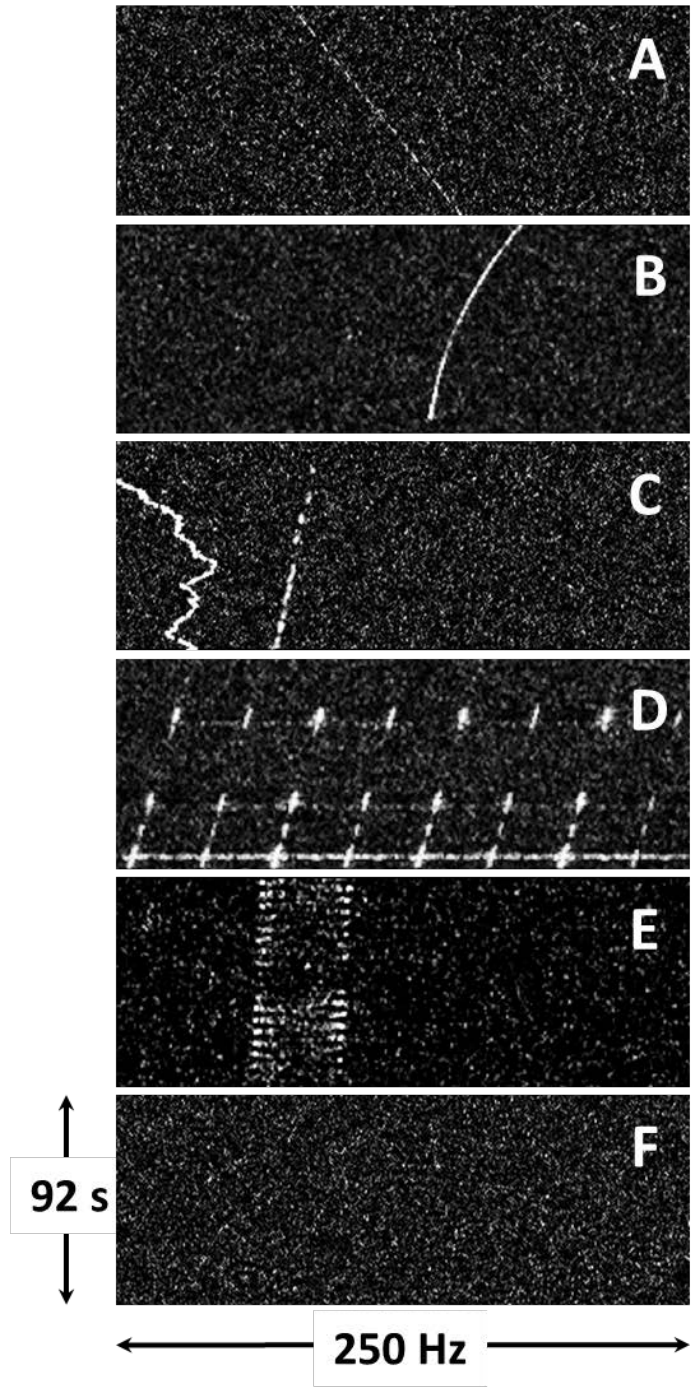

**Figure 2:** Example spectrograms from signal database. A) A narrowband drifting signal with square wave amplitude modulation; B) A narrowband signal with curvature; C) Two signals, on the left is what we call a squiggle, and to the right of that is a drifting narrowband signal; D) Impulsive signal from a wide bandwidth radar, E) A spinning satellite, and F) pure noise, well approximated by Rayleigh distribution.

For signals that exceed a threshold power, further filtration eliminates signals at frequencies with previously identified RFI. We also eliminate signals with zero frequency drift versus time, since this implies zero acceleration in the separation between the transmitter and detector. It is considered unlikely that an uncoordinated ETI transmitter could not correct for residual acceleration since it would have to match the motion of the exact telescope location on the rotating Earth. These and other filters that are initially applied before a signal is labeled as a "candidate" are described in detail in (Harp et al. 2016). When a signal is categorized as a candidate, raw voltage data are collected for a frequency range of about 600 Hz near the signal. The average candidate detection rate over the eight-year period from 2008-2016 is about 1000-10,000 candidates per hour of active observing, depending on observation frequency.

The $10^8$ collected raw voltage signals is a unique database of SETI-candidate signals that we mine to determine the prevalence and statistical distribution of signals versus 14 extrinsic variables such as telescope observing frequency, date and time of observation, apparent direction of origin, and so forth. These data are characterized using conventional statistical methods in (Harp et al. 2016). In this paper we attempt to develop new, derived quantities from the raw voltage data that enables further discrimination and sorting of these signals by signal type.

## 2.5  Database of Simulated Signals with Controlled Parametric Variables

At first, we considered the application of deep machine learning for categorization as highly speculative. We didn't know if it would work *at all*, so we decided to perform tests with a relatively uniform distribution of signals generated from parametric simulations. In the database of observed signals, a significant fraction showed power from more than one signal type at a time (e.g. from multiple transmitters at approximately the same frequency, e.g. Figure 2c). This and other uncontrolled variables might confuse the interpretation of the deep learning results. Parameterized signal simulation programs offered a means of producing a well-characterized distribution of signal types and powers.

Such a simulated set was used for the initial prototyping of deep learning categorization here. The major drawbacks of simulations are 1) the relative frequency distribution of different signal types has no relation to the distribution in real data, and 2) simulations were created for only 7 of the most commonly observed signal types; low-probability signal types were not modeled at all. This means that our quantitative deep learning results cannot be extrapolated to performance on real data. However, our primary goals were qualitative: to see if non-parametric methods would be useful.

All the simulations began with a background signal of Gaussian noise (in the complex voltage domain), which leads to a Rayleigh distribution of pixel power in the spectrograms. Examples of six classes of signals are presented in Figure 3 and were labeled as a) **narrowband**, b) narrowband with curvature (**narrowbandddrd**), c) **squiggle**, d) narrowband with square wave amplitude modulation (**squarepulsenarrowband**), e) squiggle with square wave amplitude modulation (**squigglesquarepulsednarrowband**), and f) **brightpixel**. The artificial training set comprised 20,000 simulations of each type.

The form of the brightpixel simulation is self-evident. The other signals were simulated in the time domain and can all be described by

$$s(t) = A(t) \exp(it\omega(t) + \phi) + n(t) \tag{1.1}$$

where $A(t)$ is the signal amplitude, $\omega(t)$ is the frequency, and $\phi$ is a random phase offset. The noise component was simulated by a pseudo-random Gaussian distribution of (8-bit real, 8-bit imaginary) voltage values $n(t)$, with zero mean and standard deviation of 13[2] on the full range of 0-255. The seven different signal classes were simulated with time dependent amplitude and frequency. The generic frequency function is

$$\omega(t) = \omega_0 + \omega_1 t + \omega_2 t^2 + \sum_{\tau_n = 0}^{t} B\Omega(\tau_n) \tag{1.2}$$

Where $\omega_0, \omega_1$, and $\omega_2$ are constants, $B = \text{constant}$ and $\Omega(\tau_n)$ is a uniformly sampled pseudo-random value on the domain (1,-1). The sum on the right-hand side simulates a random walk. Each simulation comprised 196608 equally-spaced time values.

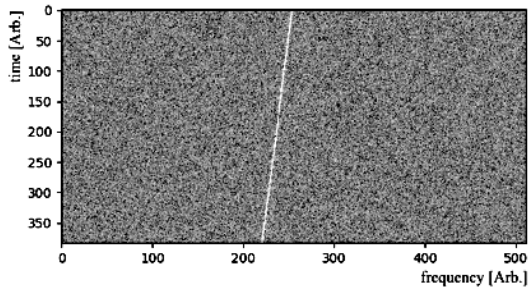
a) narrowband

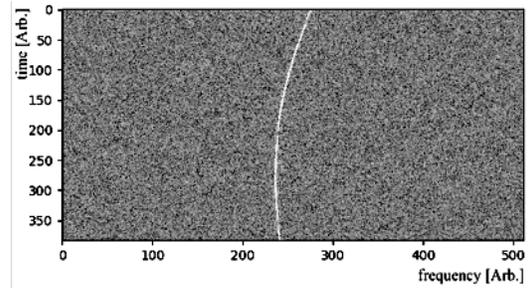
b) narrowbanddrd

---

[2] The value $\sigma = 13$ was chosen as a trade-off between having plenty of digital resolution to simulate the Gaussian noise, but also providing enough dynamic range (127 / 13 ~ 10 $\sigma$) so that rare events with high power are modeled accurately.

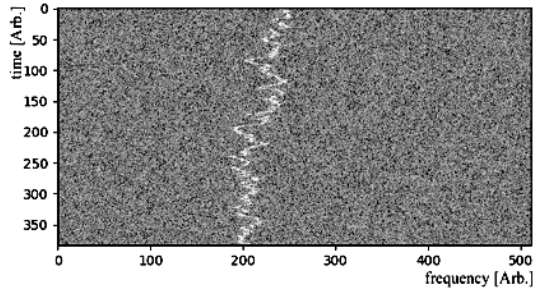
c) squiggle

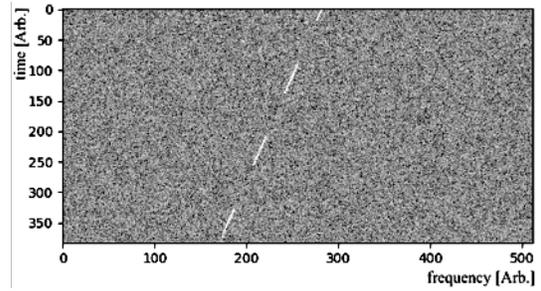
d) squarepulsenarrowband

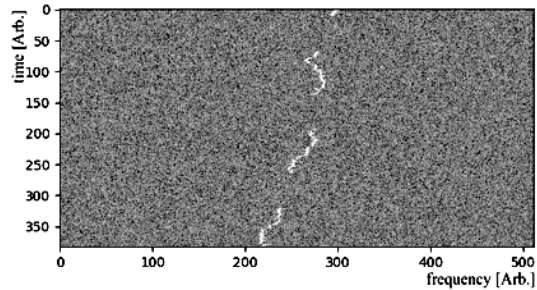
e) squigglesquarepulsednarrowband

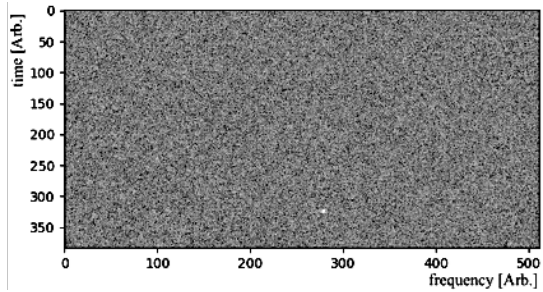
f) brightpixel

**Figure 3: Simulated spectrograms for a few example signal types. Note the similarity between some of the simulated types and the real data in Figure 2. Because the sampling rate is not specified (has no influence on spectrogram shape), the frequency and time axes have arbitrary units.**

The parameters in (1.2) are given the values by signal type according to Table 1 (top). In cases where the parameter is listed as constant (over the period of the simulation), the constant values were chosen from a uniform distribution spanning the ranges in Table 1 (bottom).

Table 1: Parameter values after Eq. (1.2) for different simulated signal types used in this work (e.g. Figure 3).

| Signal Type | $A(t)$ | $\omega_0$ | $\omega_1$ | $\omega_2$ | $B$ | noise |
|---|---|---|---|---|---|---|
| narrowband | constant | constant | constant | 0 | 0 | complex-valued pseudo-random Gaussian distribution with zero mean and $\sigma = 13$. |
| narrowbanddrd | constant | constant | constant | constant | 0 | Same |
| squiggle | constant | constant | constant | 0 | constant | Same |
| squarepulsed -- narrowband | square wave | constant | constant | 0 | 0 | Same |
| squigglesquare -- pulsenarrowband | square wave | constant | constant | 0 | constant | Same |
| noise only | 0 | 0 | 0 | 0 | 0 | Same |

| Parameter | $A_0/13$ | $\omega_0$ | $\omega_1 (10^{-6})$ | $\omega_2 (10^{-8})$ | $B$ | $T/L$ | $D$ |
|---|---|---|---|---|---|---|---|
| Range when varied | [0.05, 0.4] | [-2π/3, 2π/3] | [-7.324, 7.324] | ±[1, 8] | [0.0001, 0.005] | [0.15625, 0.46875] | [0.05, 0.9] |

A further restriction on $|\omega(t)|$ was that it was not allowed to exceed the value $\pi$ at any time, to prevent aliasing of the signal from one side of the spectrogram to the other. That is, when the computed value of the frequency Equation (1.2) fell outside of the allowed range, the signal amplitude was set to zero.

Except for the cases with square wave modulation, the value of the amplitude $A(t)$ was set to a constant value in each simulation, uniformly chosen from the range specified in Table 1 (bottom). In cases with square wave modulation, two additional parameters were chosen to define the square wave: $T/L$ specifies the pulse duration $T$ in terms of the full length of the simulation $L$, and $D$ which specifies the duty cycle.

# 3  Parametric Approaches to Categorization

Following a conventional approach for feature extraction, we developed a large number of parametric statistical indexes to extract salient features from the raw spectrograms while ignoring the noise. Examples include: 1) the mean pixel power, 2) the second, third, and fourth moments of the pixel value distribution, 3) an entropy measure (Shannon information), 4) the average of the absolute value of the pixel value less the mean value, 5) the difference between maximum and minimum pixels in the spectrogram. These indexes were computed on a) the raw spectrogram data, b) the 1D projection of the spectrogram onto the time or frequency axes, c) the numerical first derivative of the spectrogram (each pixel is replaced with the difference between the pixel and a nearest neighbor), and others.

Another feature extraction method we tried was to compute a Hough transform (Duda & Hart 1972; Hough 1959) of the 2D spectrogram, which results in a 2D representation that highlights straight line features in the data. This is very similar to, but more accurate than, the signal detection algorithm in our online signal detector called SonATA (Harp et al. 2016). By thresholding (identifying bright features in) the Hough transform we hoped to automatically characterize the frequency position and slope of drifting sinusoids in the spectrograms.

All the features described above had limited success in helping to automatically sort signal waterfall plots. For most of the statistical parameters we found a high degree of correlation between their values. The implication is that many of these statistical parameters measure "the same thing" in the spectrogram which is, essentially, the variance. The highest degree of correlation was observed between the Shannon information and all the different moments. Although dozens of parameters were computed, we discovered that they represented only a small number (2-3) of independent degrees of freedom. Similarly, analysis of the Hough transform provided discrimination between the frequencies and slopes of line-like narrowband signals, but was not especially helpful in discriminating between e.g. straight lines from squiggles (c.f. Figure 2 and Figure 3). Taken together, all these parameters can be used to sort the waterfalls crudely, but the best performance was achieved using a pattern-matching algorithm, described below.

## 3.1  *Discarding Irrelevant Information: Affine Transform*

What turned out to be the most effective derived statistic or feature for categorization used an affine transform (an implementation of the Fourier-Mellin transform, after (O'Ruanaidh & Pun 1998; Reddy & Chatterji 1996) ) to the spectrograms before comparison. The affine transform separates (and discards) translational information from the remaining information in the spectrogram. In this way, two signals that differ only in starting frequency ( $f(t=0)$ ) result in comparable images after transformation. In the same way, the transform also makes the spectrogram rotation and scale invariant (i.e. discards rotation and scale information). For completeness, we note that with the use of a real-valued spectrogram we have already discarded about half the time series information, represented as the complex phase of each spectrogram pixel.

Counterintuitively, by discarding a great deal of information irrelevant to our categorization, the resulting images were easier to categorize by type, i.e. we are more interested in in discriminating between two signals of different type than between two signals of like-type. After affine transformation, the category of a signal (e.g. narrowband, squiggle, pulse) does not depend on the starting frequency (translation), a constant frequency drift (rotation) or the time and frequency scaling of the signal. For example, all straight-line narrowband signals reduce to very similar images after transformation.

After transformation, it is possible to use the naïve sample cross-correlation coefficient between the observed spectrogram and a short list of test spectrograms to categorize signal types. The correlation coefficient of two (transformed) spectrograms designated $x$ and $y$ is computed from

$$\text{correlation coefficient} = \frac{\sum_{i=1}^{N_{pixels}} x_i \, y_i}{\left( \sum_{i=1}^{N_{pixels}} x_i^2 \sum_{j=1}^{N_{pixels}} y_j^2 \right)^{1/2}} \quad (1.3)$$

where $x_i, y_i$ are the pixel values of the signal with unknown type and the prototype signal, respectively.

Seven different prototype spectrograms, chosen from the observed data, were used in the categorization. Unfortunately, the exact spectrograms chosen as prototypes have been lost since the analysis, and we cannot display them here. But the set of prototypes was very similar to the set different signal types represented in Figures 2 and 3.

Operationally, the observed signal database was augmented with seven new columns of metadata representing the correlation coefficients of each signal with the seven prototypes. Recall that the signals were already labeled with metadata summarizing the observation conditions, including the right ascension, declination, date, time of day, center frequency, average signal power, apparent average signal slope, outside temperature, inside temperature, and so on. In all, each candidate was labelled with 21 columns of metadata.

## 3.2  Independent Component Analysis

The signal categorization amounted to a cluster analysis of all the observed signals in this 21 dimensional Hilbert space. However, not all of the different metadata columns were equally helpful in the categorization. Since we care only about identifying signal type and do not particularly care about which metadata variables have the most explanatory power, we pursued an independent component analysis (Hyvärinen et al. 2001; Stone 2004). Independent component analysis (ICA) is closely related to principle component analysis and is a method to automatically construct 21 new variables from orthogonal linear combinations of all the 21 old variables, subject to the constraint that new variables are favored when the data projection onto those variables has the most non-Gaussian distribution. The first new variable is chosen from all linear combinations to have the greatest explanatory power or discrimination in Hilbert space. If the data have the form of a hyper-cube in Hilbert space, the first independent component would be parallel to the longest axis of the hyper-cube. Then other components are determined in decreasing order of explanatory power. The last components in this expansion are often dominated by noise (have almost zero explanatory power) especially when the original variables are correlated (i.e. not orthogonal to one another).

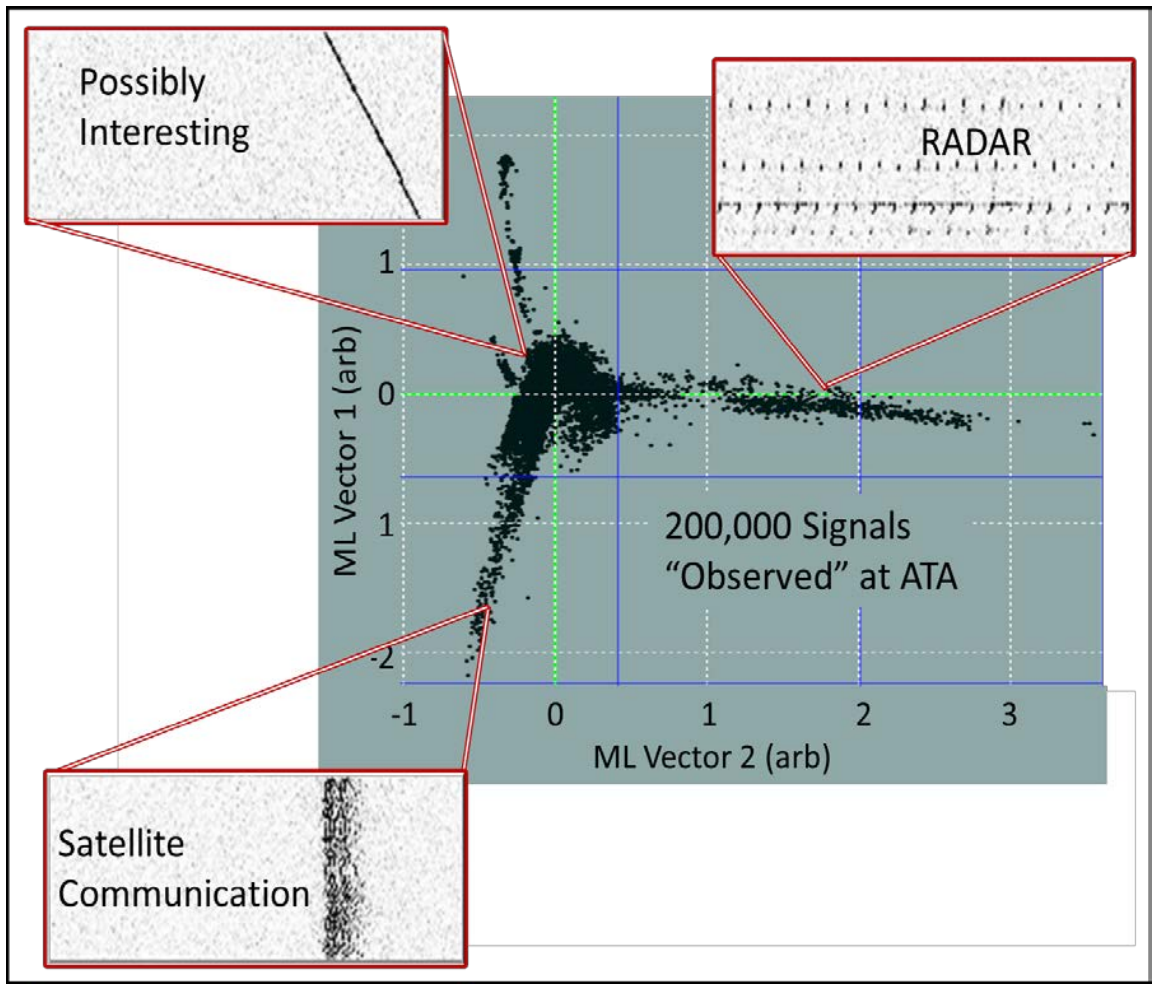

**Figure 4: Two hundred thousand spectrograms are subjected to independent component analysis (IDA) of all the metadata variables described here, including intrinsic variables (observing frequency, time and date, etc.) and derived variables such as the post-affine transform correlation coefficient with seven different test spectrograms analogous to those in Figures 2 and 3. In the figure, each black dot represents a spectrogram. The**

plotting axes are the first and third independent components. Exemplary spectrograms from different parts of the figure show how ICA sorts the spectrograms by type. The large blob near the origin contains spectrograms that appear, to the eye, to contain only noise.

The goal of ICA is dimensional reduction. Because the computational complexity of a cluster analysis varies exponentially with the number of independent variables, even a reduction of 1 dimension can greatly simplify the analysis. Consider two variables that are highly correlated, such as time of day and outside temperature (because it is usually cooler at night). If these two original variables were perfectly correlated, then the spectrogram data are restricted to a $N-1$ dimensional hyperplane in $N$-dimensional Hilbert space. By proper choice of linear combinations, we can make one of the new variables orthogonal to this hyperplane. This orthogonal variable can then be discarded since it would have no explanatory power in modeling the data.

The way that ICA aids signal categorization is typified in Figure 4. Here, each point represents a spectrogram and we plot the data in two dimensions along the first and third independent components. Although there is no guarantee that discrimination along these axes corresponds to our desired typology, in practice it does.

### 3.3   Prescription for a Signal Categorizer Based on ICA

To build a categorizer based on the components in Figure 4, one can simply write down the expressions for the most helpful independent components in terms of the input variables. The categorizer then projects the metadata of an unknown signal onto these components and labels are assigned according to where the signal falls in a plot like Figure 4. The categorizer assigns labels derived from a cluster analysis of the 200,000 observed signals.

Given the excellent discrimination observed in Figure 4, we are confident that a useful categorizer could be derived from an ICA analysis trained with a database of real observed signals of all types. The initial ICA is extremely computationally expensive and would have been unfeasible without the power of the supercomputer center at NASA Ames. However, after the ICA is complete, categorization requires only the computation of dot products of the unknown signal's metadata with the first few independent components. The latter is a much simpler calculation requiring far less computer resources than the initial ICA. Thus we arrive at a categorizer that is relatively fast.

One step we have skipped over until now is the hand-assignment of different signal class labels to different parts of the space spanned by the independent components. This might be a laborious effort, but it was dramatically simplified thanks to NASA's hyperwall display. The NASA supercomputer resources were available to us for only a limited time. This is the reason that the entire process could not be repeated even after loss of the exact prototype spectrograms used for computing correlation coefficients (see section 2.1).

## 3.4 Summary of Parametric Analyses

We have attempted computation of a large variety of features (parameter values) assigned to or derived from the spectrogram data. While many of these parameters had a small but significant explanatory power for signal categorization, by far the most successful parameters were obtained by comparison of unknown spectrograms with a small number of prototype spectrograms hand-selected from the data. This comparison was made by first applying an affine transform to both the unknown and prototype spectrograms to eliminate uninteresting translational, rotational, scale, and phase dependent information embedded in the data. A numerical value quantifying each comparison was computed from the cross-correlation coefficient of unknown signals with the prototypes.

Once we had found a set of derived parameters with high explanatory power, an independent component analysis of all metadata including these parameters was performed. Qualitatively, we observed that only a small number (3-4) independent components were enough for useful signal labeling. We described a prescription by which a fast, useful categorizer could be constructed using these independent components. This is a prototype study, so we were satisfied with demonstrating a recipe for building a classifier. The development of an actual classifier that can be applied to spectrogram data in near-real-time is a task left for future work.

## 4 Non-parametric Categorization

We frame the challenge of classifying radio signal time series of complex amplitudes as an image recognition task. This allows us to apply, wholesale, convolutional neural network image labeling algorithms that have been so effective in recent research, beginning with the breakthrough study of (Krizhevsky A et al. 2012). Since then, a variety of network architectures have been developed for image classification, gradually improving performance to superhuman levels. In the most famous "John Henry" example, Andrej Karpathy competed against a neural network in the task of image classification in the ImageNet Large Scale Visual Recognition Challenge (Karpathy 2014). Karpathy found that a trained human classifier can achieve an error rate of 5.3% in this task. In 2014, Karpathy outperformed the best neural network classifier available, but by 2017, neural networks routinely were able to achieve error rates below 5% (Gershgorn 2017).

## 4.1 Exploration of CNN Architectures

Deep learning is an active area of research in supervised machine learning tasks. For image recognition, convolutional neural networks (CNNs) are at the core of these advances, and the gains in state-of-the-art performance demonstrated by residual networks (ResNets) is a go-to example (He et al. 2016). We tested multiple convolutional neural network architectures in this classification task using only the spectrogram (power) images as input features.

There is no need to reinvent the network topology; instead we use tried-and-true network configurations known to be excellent image classifiers, beginning with the 2015 ImageNet winner created by the ResNet team. Augmentations and further improvements to the ResNet architecture have been presented since, so we tested these in a benchmark against ResNet and each other to determine which performed best on the simulated signal dataset. All these competing networks were able to show state-of-the-art performance on this task with similar accuracies. As this work was done in the context of a competition, focus was put on finding the CNN with the absolute highest performance.

### 4.1.1 Residual Network (ResNet)

ResNet introduces the residual connection (c.f. Figure 5) between convolutional layers in a very deep convolutional neural network in order to combat the loss of signal backpropagation. Previously, very deep networks tended not to train successfully, but this skip connection that bypasses a layer's nonlinearity allows gradients to backpropagate further in the network and allow for deeper, more expressive networks. Hence, we experimented with ResNets up to the limit of computational constraints for these medium resolution images.

### 4.1.2 Wide Residual Network (WRN)

The development of wide residual networks was motivated by observations that increasing the depth of ResNet provided diminishing returns on network performance improvement, and offered shallower convolutional networks with more convolutional filters at each layer as an alternative (Zagoruyko & Komodakis 2016a). Additionally, the authors change the order of convolution, batch normalization, and activation, and add dropout to tune their architecture to train faster and perform competitively with deeper ResNet models. In our experiments, we used the variations with highest performance in previous work; specifically, using the same 3x3 basic convolutional block (as opposed to a bottleneck block), and a dropout rate of 0.3.

### 4.1.3 Densely Connected Residual Network (DenseNet)

DenseNet extends the idea of residual connections by adding connections not only between consecutive convolutional layers, but also between all subsequent convolutional layers (Huang et al. 2017). The dense residual block allows the gradient signal to skip more layers, and which more closely ties the loss function to earlier layers of the network. The additional skip connections also are thought to encourage feature reuse by sending signal from multiple convolutional layers to subsequent layers, leading to more expressive power in a more compact network.

### 4.1.4 Dual Path Network (DPN)

Finally, we experimented with dual path networks, which integrate both residual networks and densely connected residual networks to realize the advantages of each, while sharing weights to maintain a reasonable model complexity (Chen et al. 2017).

## 4.2 Comparison of Models

We trained several differently-sized models for the ResNet, WRN, DenseNet, and DPN architectures. The accuracy of a learned classifier depends on this size because less complex networks cannot necessarily express as complex a pattern as a larger network, but the larger networks do not necessarily succeed (converge) in learning the patterns we wish to express. Table 2 lists the models trained for each architecture type, with the best performing model shown in bold.

Table 2: Comparison of model architectures. Accuracy is measured with a cross-validation subset of the full training set. The number of parameters is given for the best-performing model (bold) of each type.

| Architecture | Models Tested | Accuracy | Parameters |
|---|---|---|---|
| **ResNet** | ResNet-18, ResNet-50, **ResNet-101** ResNet-152, and ResNet-203 | 94.99 | 42.6M |
| **Wide ResNet** | **WRN-34-2**, WRN-16-8, and WRN-28-10 | 95.77 | 1.9M |
| **DenseNet** | DenseNet-161, **DenseNet-201** | 94.80 | 18.3M |
| **Dual Path Network** | **DPN-92**, DPN-98, and DPN-131 | 95.08 | 35.1M |

The accuracies included in Table 2 are on the same train-validate split for all architectures. The performances of the four best models are comparable, with the WRN-34-2 (Figure 6) only slightly outperforming the rest. However, a very significant basis for choosing the WRN-34-2 over the other models is the size of the model. The WRN-34-2 contains only 1.9M parameters (also called weights). In comparison, the second smallest model, the DenseNet-201, is nearly ten times as large. The ResNet-101 is the largest, with over twenty times as many parameters. For the purposes of quick training and inference on-site at the ATA, it is advantageous to have a smaller number of products to compute.

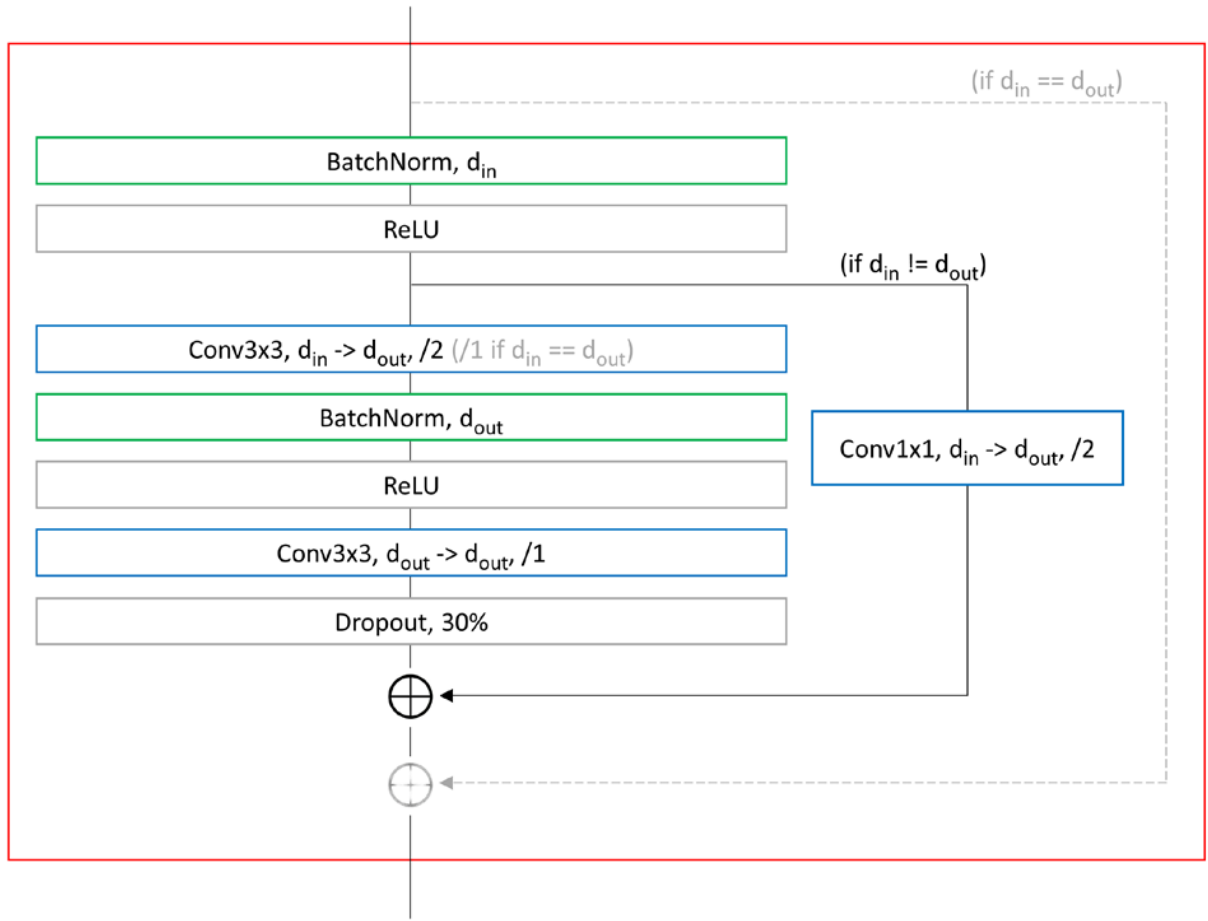

Figure 5: Basic Block: two 3x3 2D convolutions with batch normalization, ReLU nonlinearity, and a residual connection.

## 4.3   Wide ResNet Ensemble

The comparison of these four network architectures showed similar and encouraging results, and the edge that WRN-34-2 had over the others led us to continue working to improve its performance. A common practice to increase generalizability of a machine learning model that threatens to overfit on training data is to create an ensemble of models. For this task, ensemble averaging was used. Five WRN-34-2 models were trained on different subsets of the training data but with the same hyperparameters and training strategy. While the comparative study of architectures used the same four of five folds for training and the fifth for validation, the ensemble members trained on the five distinct four-fold subsets of data, as with k-fold validation for k = 5.

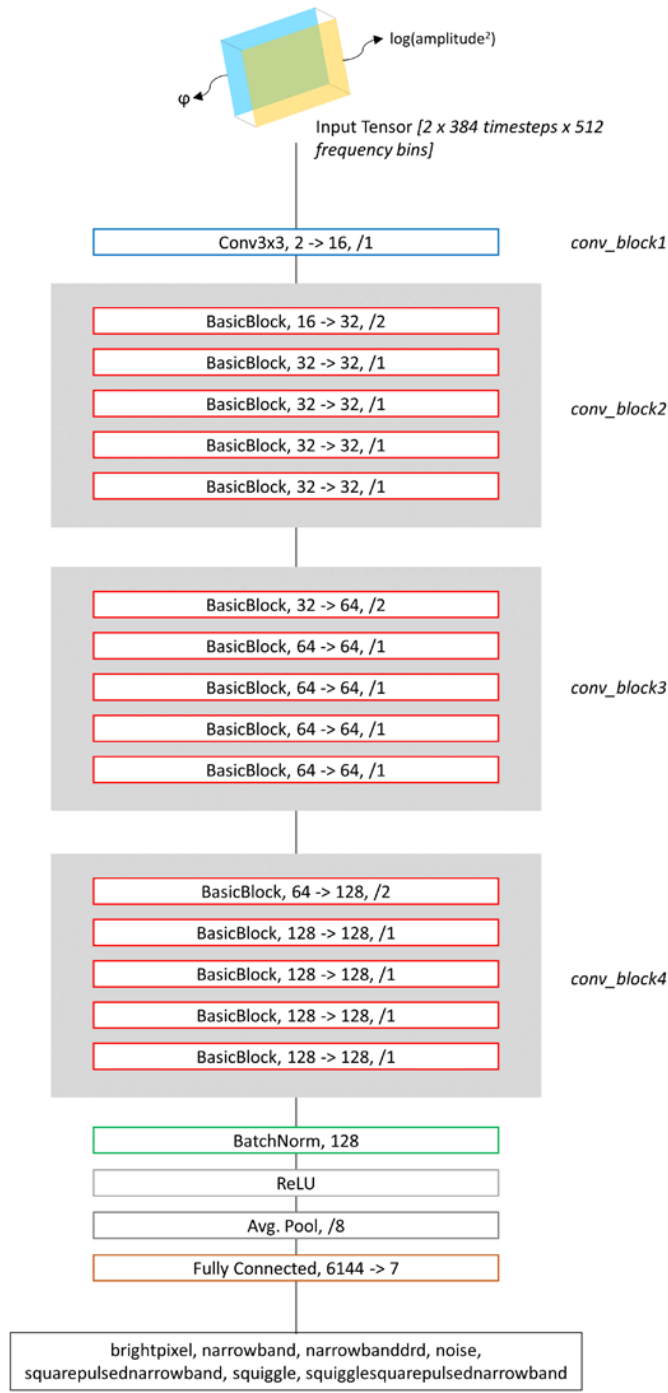

Figure 6: Complete WRN-34-2: a series of Basic Blocks grouped into three sections of increasing number of features planes and decreasing image resolution.

To evaluate the ensemble model, each of the five-member models outputs its softmax predictions and the average of these scores is taken as the final score. In final tests, this ensemble model was evaluated on new test data provided only near the end of the code challenge, which yielded an accuracy of 94.99%.

## 5    Final Model Performance

### 5.1    Classification Report

The model with the best validation set accuracy, the WRN-34-2 using a 5-fold averaging, was then tested using a separate test set withheld during the training phase of all models. As this work was performed in the context of an online code challenge, the other models described in the previous section were not tested with this test data set. Table 3 and Table 4 display the performance of the WRN model on the test set of 2496 simulations.

Clearly the largest source of uncertainty was distinguishing brightpixel signals from noise. As we'll see in the next section, this was due to very low-amplitude brightpixel signals, as one would intuitively expect.

Table 3: Model Performance Scores. The number of simulations of each class in the test set is given by N.

|  | N | precision | recall | $F1 = 2 \dfrac{\text{precision} \times \text{recall}}{\text{precision} + \text{recall}}$ |
|---|---|---|---|---|
| brightpixel | 385 | 0.991 | 0.857 | 0.919 |
| narrowband | 355 | 0.994 | 0.944 | 0.968 |
| narrowbanddrd | 348 | 0.969 | 0.977 | 0.973 |
| Noise | 368 | 0.785 | 0.995 | 0.877 |
| squarepulsednarrowband | 385 | 0.975 | 0.925 | 0.949 |
| squiggle | 322 | 1.000 | 0.997 | 0.998 |
| squigglesquarepulsednarrowband | 332 | 1.000 | 0.970 | 0.984 |

Table 4: Confusion Matrix. Actual classifications are along each row and predicted classification counts are along the columns. For example, there were 322 squiggle signals (Table 3). Of those, 321 were predicted to be squiggles and one was predicted to be noise. Almost all misclassifications appear in the noise column. Labels have been abbreviated: bp for brightpixel, nb for narrowband, drd for narrowbanddrd, no for noise, sqPnb for squarepulsednarrowband, sgl for squiggle, and sglsqPnb for squigglesquarepulsednarrowband.

| Actual / Predicted | Bright pixel | narrowband | drd | noise | sqnb | squiggle | sglsqPnb |
|---|---|---|---|---|---|---|---|
| bp | 330 | 0 | 0 | 55 | 0 | 0 | 0 |
| nb | 0 | 335 | 10 | 5 | 5 | 0 | 0 |
| drd | 0 | 0 | 340 | 7 | 1 | 0 | 0 |
| no | 1 | 0 | 0 | 366 | 1 | 0 | 0 |
| sqPnb | 2 | 2 | 1 | 24 | 356 | 0 | 0 |
| sgl | 0 | 0 | 0 | 1 | 0 | 321 | 0 |
| sglsqPnb | 0 | 0 | 0 | 8 | 2 | 0 | 322 |

## 5.2 Model Performance Characteristics

In order to briefly explore the performance characteristics of this trained model in a controlled way, 14 new sets of test data were generated, each with 250 signals of each class. For each set, the signals were simulated with a fixed signal amplitude, A/13.0 = 0.008, 0.01, 0.02, 0.04, 0.05, 0.06, 0.07, 0.08, 0.09, 0.1, 0.12, 0.16, 0.2, or 0.4. Many of these signal amplitudes, it should be noted, are below the amplitudes of signals found in the training data set, allowing us to explore the model performance slightly outside of range of signal amplitudes on which it was trained. All other parameters of the test data, however, remained consistent with the training data.

For each of these test sets, we performed inference and recorded the model's F1 score (Figure 7a), multinomial cross entropy loss, and classification accuracy (Figure 7b). The model performs as expected. Signals with smaller amplitudes were more difficult to classify and tend to be classified as noise. Also, the model does not appear to have any classification power with signal amplitudes below the trained signal amplitude space.

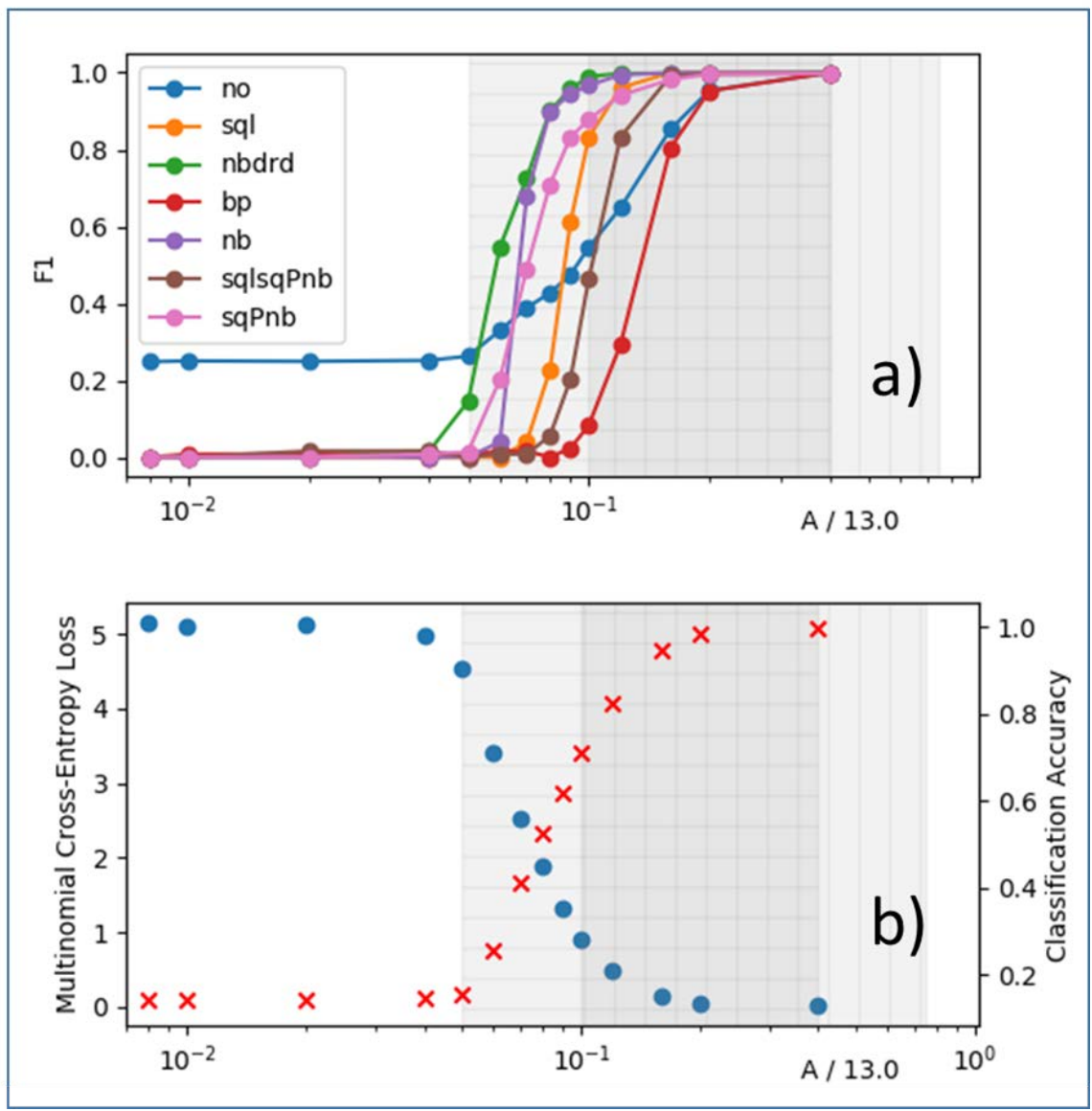

Figure 7: F1 scores, multinomial cross-entropy loss scores, and classification accuracy versus the signal amplitude (A/13.0). The horizontally hatched region represents the signal amplitudes in the training data with amplitudes in the range [0.05, 0.4], while the vertically hatched region spans the amplitudes from [0.1, 0.75]. Refer to Table 4 for the signal classes that were simulated in these regions.

The onset rise of each class's F1 score Figure 7a can be intuitively explained by considering the average amount of power per pixel in each signal, akin to the apparent brightness of the signal in the spectrogram. The brightpixel signals have lowest total power, overall. These signals have zero amplitude for most of the simulation except for a brief moment at a particular frequency. As such, the model struggles the most to recognize these signals. The squiggle simulations with non-zero B (squiggle and squigglesquarepulsednarrowband) also have a reduced apparent brightness in the signal portions spectrogram. The stochastic fluctuations of these signals result in the power being spread across a larger frequency bandwidth than with narrowband, even though the overall powers are comparable. In the pixels containing significant signal power, the squiggles have a reduced average brightness than for narrowband. The narrowband and narrowbanddrd have the earliest onset of significant F1 scores, followed by squarepulsednarrowband, consistent with the argument that the brighter signal types are more easily recognizable.

## 6 Discussion

### 6.1 Comparing Parametric and Non-parametric Methods

The conventional appellation "non-parametric" referring to CNNs is somewhat of a misnomer, since CNN models obviously have adjustable parameters (i.e. Table 2). The important difference between the models described in sections 3 and 4 is in the definition of feature extraction and the model's functional form. In section 3, the bottleneck in finding a solution is in trying many different feature extraction methods to find ones with good predictive power, and the model has a very simple multi-linear form.

Thanks to the universal approximation theorem (Hornik 1991) the NN has a very large space or set of functional forms to choose from, even non-analytic and non-continuous functions. The researcher is freed from having to choose the form of the model since the functional form is essentially another parameter in the fitting process.[3] The down side is that adequately fitting all these new parameters requires a much large training data set than a hand-crafted model with minimal parameterization. This is the main bottleneck associated with developing good NN models; a dramatically larger number of training operations are required to obtain comparable performance. This is why neural networks have only recently shown great success though they were invented in 1958 (Rosenblatt 1958). Only now do we have sufficient computing power to train a complex neural network.

---

[3] There is no guarantee that the CNN will converge on the optimal functional forms and parameter values. In fact, the opposite is true. It is understood that CNNs work because there are many combinations of models and parameter values that give nearly optimal results. If model parameters are likened to atoms in a multicrystalline solid, there are many different meta-stable configurations in which the system can become trapped, all of which have an energy very close to the minimum. In solids, the time required for the configuration to settle into a truly stable state with minimum energy is so long that it is rarely seen in nature. Similarly, two well-trained CNNs giving almost maximum performance may have no structures in common with one another.

Neural networks have the advantage that we need not be concerned with feature extraction either. In principle, one can simply dump all available information into the model input, regardless of whether it is relevant or not. Given enough time, the NN will learn its own feature extraction methods and learn to ignore irrelevant data. However, if only relevant information is provided to the NN at training time, then the time to converge will be greatly shortened.

For this reason, we suggest a combination of parametric and non-parametric methods will lead to the best model. For example, the training time may be much improved by preprocessing spectrogram data with an affine transform before feeding it into the CNN. Given a fixed computational cost, shorter training times allows e.g. for the development of multiple parallel CNNs whose combined predictions are better than any single CNN. Another way to combine the benefits of both parametric and non-parametric methods is to run their associated categorizers in parallel and then use a combined prediction at run time.

But what about the spectral phase information we discarded at the beginning, when we formed the spectrograms from the complex-valued Fourier transforms? Additional work, not shown here, has tested the hypothesis that relevant signal information is readily accessible in raw phase images. Based on the best performing NN configuration, the network input was modified to accept both the spectrogram and phase image. This network was then trained using the same protocol as for the CNNs described above. After a comparable number of training iterations, this network did not perform significantly better than those using the spectrogram alone. We tentatively conclude that, at least for the purposes of categorization, the phase information does not contribute much information.

### 6.2   Sensitivity to Weak Signals

Our simulated training dataset contained a substantial number of weak signals with amplitudes as low as 5% the noise power. It was somewhat disappointing that after training to convergence, the CNNs were blind to the weakest signals (c.f. misclassifications in Table 4 and turn-on amplitude in Figure 7). A hint as to the reason why is found in Figure 7, where spectrograms with the signal power was concentrated into the fewest number of pixels were most easily characterized. A similar effect was observed in the parametric studies of section 3 (results not shown).

This may be due partly to the mismatch between the signal to noise ratio in our spectrograms relative to typical photographs, as mentioned in section 2.2. In fact, recent research comparing human and neural network classifiers has found that humans are much more successful than NNs in the presence of moderate noise (Geirhos et al. 2017). Another hint is found in the phenomenon of *adversarial examples.* Szegedy et al. discovered that well-trained neural networks can be spoofed, that is, caused to misclassify by applying small image perturbations hardly perceptible to the human eye (Kurakin et al. 2017; Paloncýová et al. 2012). This phenomenon is not yet well understood, but it indicates that neural networks are highly sensitive to image information that is less important to the human perceptual system. In some examples, the perturbations look almost like additive random noise (Kurakin et al. 2017). This is consistent with the observation that neural network categorizers are especially fragile to additive noise.

The model's performance falloff just at the lower-bound of the signal amplitudes (section 5.2) leads to a question: to what small signal to noise ratio can we train a model of the same network architecture and still retain robust classification accuracy? That is, if we were to construct new training data with smaller signal amplitudes and retrain new models, how small in amplitude can we go before the models fail to accurately classify signals?

Our speculation on this topic is as follows: the fact that the noise component is uncorrelated between images suggests that the noise spans a very high-dimensional space in the information domain. Recognizing noise, therefore, would require a more complex neural network and many more training examples than a network that classifies simpler features.

## 6.3    Unanswered Questions and Directions for Further Study

### 6.3.1    Multiple Signals

Real observations from the ATA do not always contain just one signal or signal type in the spectrogram. In very recent work (Rankawat & Harp 2018) the network was extended to deal with multiple, unrelated signals.

### 6.3.2    Improvements to Simulations

Future work building on the results here could focus on several aspects. More signal types could be added to training and test sets. For example, a common signal type not included in our repertoire is a short burst of power over all frequencies sometimes identified with radar pulses.

The noise component of the signal $n(t)$, could be more realistic. For this study, we settled on a very simple Gaussian white noise model. In a previous version of the training data set, however, we used real observations from the Sun as the background component, $n(t)$. On the timescales important here, the Sun has approximately stationary power at all frequencies, and because it emits very high flux, it dominates all the undesirable radio frequency interference that would normally appear. Also, captured data provide a better model for the nonlinear bandpass filter (pink noise) effected in a real telescope.

### 6.3.3    Unsupervised Learning

Given our very large database of archived SETI signals, an unsupervised learning approach might be useful. T-distributed Stochastic Neighborhood Embedding (t-SNE) is a method for visualizing high-dimensional datasets, since it is a well-balanced dimensionality reduction algorithm that requires no labels yet reveals latent structure in many types of data. One such attempt is reported in (Luus 2017), who used an autoencoder and t-SNE clustering applied to real data observed with ATA data from 2013 to 2015. While much of the clustering was unsupervised, Luus used partial labeling as an aid for attaching meaning to identified clusters.

### 6.3.4 Predicting Future Signal Behavior

The line fitting algorithm currently in use at the ATA estimates a linear fit to signals in observed spectrograms, with parameters including the signal power, initial signal frequency, and drift rate. This regression is important because it allows for prediction of the future signal frequency, which is needed to identify the same signal in a subsequent observation. Such parameterization could be extended to more signal types; once the signal type has been identified, we could choose the correct signal model, fit to the model, and extrapolate from there. This is not a new idea, and the SETI literature contains many examples of different parametric models. But apart from (potentially dispersed) wideband pulses (Von Korff et al. 2013), none of these models have been implemented in large scale searches. The computational requirements of complex parametric models have always been too high a price to pay, when compared with expanding the existing narrowband models.

More interestingly, we can repose the problem as the comparison of two raw spectrograms, asking the question whether the later spectrogram contains a signal that follows from the first. An ordinary CNN could be trained with pairs of real or simulated spectrograms taken at different times. Some pairs would show spectrograms a continuous time series. Other pairs would contain images of unrelated signals or noise. Such a CNN, when trained, might be able to identify cases where the signal in the second spectrogram is related to that in the first. This CNN would be signal-agnostic and be used to correlate any of the signal types studied here.

A third approach uses a recursive neural network to predicts a signal's future shape based on a time-series describing how the signal develops in time (Zhang et al. 2018b). In that paper, they also use an adversarial approach to unsupervised network training. Unsupervised training allows for learning from exceptionally large datasets that would take too long to label by hand. This approach seems quite promising.

# 7 Conclusions

This paper explores the adaptation of machine vision algorithms for the categorization of artificial signals commonly observed in a SETI search. The one-dimensional voltage time series known to contain signals was rendered in two-dimensional spectrogram images, and methods used for the classification of photographic images were prototyped using these spectrograms. We began with a large archive of observed non-natural signals gathered over eight years of observations at the ATA. We prototyped many algorithms that might be used to classify spectrograms obtained from those signals.

Because of the relatively low signal to noise level in observed spectrograms, it was not obvious whether machine vision algorithms would be effective, especially considering the documented failure of some algorithms in the presence of moderate noise (Geirhos et al. 2017). We found that machine vision categorization was highly effective when the signal to noise ratio was relatively high.

We tested many statistical measures or features extracted from spectrograms that might be useful for classification. Most of these features showed limited categorization power. The most effective features were extracted in a two-step process: Firstly, the spectrograms were treated with an affine transform that produced images invariant to translation, rotation, and scaling operations on the original spectrogram. Secondly, a simple cross correlation coefficient was computed in comparison of the unknown signal with the transforms of a small set of prototype signals selected in advance. The resultant correlation measures had a great deal of predictive power for distinguishing different classes of unknown signals. Simply put, by throwing away phase[4], translational, rotational, and scale information, simple correlation-based similarity comparisons could be used as a powerful aid to classification.

This work also demonstrates the potential usefulness of applying contemporary convolutional neural networks to spectrogram analysis. We report the adaptation of a convolutional neural network, specifically a wide residual network, to the problem of signal discovery and classification relevant to SETI. We find that by treating spectrograms as if they were images, we can train an image classifying network and achieve very good results.

The methods described here do not compete with a finely-honed conventional algorithm in terms of signal to noise ratio. We empirically estimate the minimum signal to noise ratio (SNR) for the CNN is about 10x greater than the minimum SNR for the standard line fitting algorithm used in our workhorse SETI detector.

But even with this restriction, the flexibility of CNNs to categorize a wide range of signal types could be used to great benefit in a near-real-time SETI observing system. We envision the categorizer to be placed after the first stage of signal detection performed by the existing SonATA narrowband detector. Thus, the categorizer sees only a small subset of all the spectrograms tested through the system. The categorizer predictions are combined with other metadata and become new columns in the archive database. Then the follow-up logic for interesting signals can be informed by the category. For example, the SonATA detector is optimized for very narrowband straight-line signals. Sometimes this detector finds squiggles, but we have no confidence that a squiggle found in one spectrogram will be reliably found in a subsequent spectrogram of the same signal, even if the signal is real. Based on this information, we may choose to not follow-up on squiggle signals (until a better detector is built) thereby saving observing time for the study of other targets and frequency ranges.

---

[4] The step of discarding phase information is actually performed during the computation of the spectrogram.

Compared to the conventional algorithm, the machine vision approaches lack the parameterization of signals necessary for extrapolation to later times, which allows us to judge whether the same signal source is observed at two different times. Future modifications of the regression schemes or new neural networks (Zhang et al. 2018b) can be used to determine whether two spectrograms contain signals from the same source.

# 8 Acknowledgements

We'd like to thank Galvanize, Skymind, Nimbix, and The SETI League for their financial contribution and to the hard work put in by many employees of those organizations that ensured a successful code challenge. Thanks to IBM for providing significant compute and data storage. We acknowledge the helpful suggestions from Francois Luus of IBM Research South Africa.